\documentclass[aps,prb,preprint,superscriptaddress,groupeaddress]{revtex4}
\usepackage{graphicx}
\usepackage{amsmath}
\usepackage{mathrsfs}
\usepackage{accents}
\begin{document}

\title{Realization and transport investigation of a single layer-twisted bilayer graphene junction}

    \author {Dingran Rui}
    \affiliation{Bejing Key Laboratory of Quantum Devices, Key Laboratory for the Physics and Chemistry of Nanodevices and Department of Electronics, Peking University, Beijing 100871, China}
    \author {Luzhao Sun}
    \affiliation{Center for Nanochemistry, Beijing Science and Engineering Center for Nanocarbons, Beijing National Laboratory for Molecular Sciences, College of Chemistry and Molecular Engineering, Peking University, Beijing 100871, P. R. China}
    \affiliation{Academy for Advanced Interdisciplinary Studies, Peking University, Beijing 100871, China}
    \author {N. Kang}
    \email[Corresponding author: ]{nkang@pku.edu.cn}
    \affiliation{Bejing Key Laboratory of Quantum Devices, Key Laboratory for the Physics and Chemistry of Nanodevices and Department of Electronics, Peking University, Beijing 100871, China}
    \author {J. Y. Li}
    \affiliation{Bejing Key Laboratory of Quantum Devices, Key Laboratory for the Physics and Chemistry of Nanodevices and Department of Electronics, Peking University, Beijing 100871, China}
    \author {Li Lin}
    \affiliation{Center for Nanochemistry, Beijing Science and Engineering Center for Nanocarbons, Beijing National Laboratory for Molecular Sciences, College of Chemistry and Molecular Engineering, Peking University, Beijing 100871, P. R. China}
    \author {Hailin Peng}
    \affiliation{Center for Nanochemistry, Beijing Science and Engineering Center for Nanocarbons, Beijing National Laboratory for Molecular Sciences, College of Chemistry and
    Molecular Engineering, Peking University, Beijing 100871, P. R. China}
    \author {Zhongfan Liu}
    \affiliation{Center for Nanochemistry, Beijing Science and Engineering Center for Nanocarbons, Beijing National Laboratory for Molecular Sciences, College of Chemistry and
    Molecular Engineering, Peking University, Beijing 100871, P. R. China}
    \author {H. Q. Xu}
    \email[Corresponding author: ]{hqxu@pku.edu.cn}
    \affiliation{Bejing Key Laboratory of Quantum Devices, Key Laboratory for the Physics and Chemistry of Nanodevices and Department of Electronics, Peking University, Beijing 100871, China}
    \affiliation{Beijing Academy of Quantum Information Sciences, Beijing 100193, China}
    \date{\today}

\begin{abstract}
We report on low-temperature transport study of a single layer graphene (SLG)-twisted bilayer graphene (tBLG) junction device. The  atomically sharp SLG-tBLG junction in the device is grown by chemical vapor deposition and the device is fabricated in a Hall-bar configuration on Si/SiO$_2$ substrate. The longitudinal resistances across the SLG-tBLG junction (cross-junction resistances) on the two sides of the Hall bar and the Hall resistances of SLG and tBLG in the device are measured. In the quantum Hall regime, the measurements show that the measured cross-junction resistances exhibit a series of new quantized plateaus and the appearance of these resistance plateaus can be attributed to the presence of the well-defined edge-channel transport along the SLG-tBLG junction interface. The measurements also show that the difference between the cross-junction resistances measured on the two sides of the Hall-bar provides a sensitive measure to the edge channel transport characteristics in the two graphene layers that constitute the SLG-tBLG junction and to degeneracy lifting of the Landau levels in the tBLG layer. Temperature dependent measurements of the cross-junction resistance in the quantum Hall regime are also carried out and the influence of the transverse transport of the bulk Landau levels on the edge channel transport along the SLG-tBLG junction interface are extracted.
These results enrich the understanding of the charge transport across interfaces in graphene hybrid structures and open up new  opportunities for probing exotic quantum phenomena in graphene devices.
\end{abstract}

\maketitle

The physics of graphene, a two-dimentional material with honeycomb lattice, has been intensively investigated over the past decade\cite{Novoselov2004,Novoselov2005,Novoselov2006,ZhangY2005}. Unlike massive charge carriers in conventional semiconductors, carriers in single layer graphene (SLG) exhibit a linear energy-momentum dispersion relation and behave as chiral massless particles.
In the case of bilayer graphene, stacking two graphene layers on top of each other can cause a drastic change in the electronic band structure, inducing new intriguing quantum states and physics properties.
In ubiquitous AB-stacked bilayer graphene, the massive quasiparticles exhibit a quadratic dispersion near the Dirac point\cite{Novoselov2006}. In twisted bilayer graphene (tBLG) with a sufficiently large rotational angle between the two layers, the energy bands of the two layers retain their linear energy dispersions at the Dirac points\cite{DosSantos2007} and the Dirac cones of the two layers intersect, leading to the formations of saddle points in reciprocal space and van Hove singularities in the density of states\cite{LiG2010,YanW2012}. The superposition of the lattices of the top and bottom layers also gives rise to the emerging moir\'e pattern, leading to the formation of Hofstadter butterfly energy spectra \cite{Bistritzer2011,KimK2017}. In tBLG with a small twisted angle, the interlayer coupling strongly depends on the twisted angle \cite{KimY2013,KorenE2016}, making it possible to host strong correlated quantum states.
Very recently, both superconducting and Mott-like insulating states have been observed in tBLG near the ``magic angle",\cite{Cao2018,CaoYuan2018,Yankowitz2019,Lu2019} opening up a new way of exploring strong correlated systems.

In the presence of a relatively large magnetic field, graphene exhibits unusual quantum Hall effect, along with the chiral edge states. As for SLG, the quantized Hall resistance $R_{xy}$ follows $1/R_{xy}$=4e$^2/h(n+1/2)$ with integer number $n$.\cite{ZhangY2005,Novoselov2005} For the case of tBLG with a large twisted angle, the Hall plateau resistance follows a different progression $1/R_{xy}$=8e$^2/h(n+1/2)$ due to the eightfold degeneracy of the Landau Levels\cite{DeGailR2011}, originating from the spin, valley, and layer degrees of freedom\cite{SanchezYamagishiJD2012}.
In the quantum Hall regime, one-dimensional chiral edge states form at the boundaries.
Recently, there is growing interest in the transport studies of equilibration and interference of the edge states using graphene hybrid structures\cite{Williams2007,Klimov2015,SchmidtH2013}.
Nevertheless, the study of quantum Hall states in tBLG has so far mainly been performed on homogenous samples\cite{Smet2011,SanchezYamagishiJD2012}. The transport properties of SLG-tBLG hybrid structures have seldom been explored.

Several methods have been employed for producing tBLG, including epitaxy on the C surfaces of SiC\cite{HassJ2008}, artificial stacking by dry-transfer technique\cite{SanchezYamagishiJD2012,ChaeDH2012} and chemical vapor deposition (CVD)\cite{ReinaA2008,LiaoL2015}.
Recently, we have achieved high-quality tBLG by means of CVD\cite{LiaoL2015, YinJ2016}.
These CVD-grown tBLG samples contain naturally formed SLG-tBLG junctions with an atomically sharp interface. Considering the difference in edge channels formed in single layer and twisted bilayer graphene in the quantum Hall regime, these SLG-tBLG junctions offer an intriguing platform to investigate the edge state coupling in graphene heterostructures. It is also interesting to study how the junction interface impedes the electron transport across the junction\cite{Giannazzo2012}.

In this work, we report on an experimental study of the transport properties of a SLG-tBLG junction device. Our high quality SLG-tBLG junction is synthesized by CVD and the device was made in a Hall-bar configuration on a Si/SiO$_2$ substrate. Low-temperature transport measurements show that the tBLG region in the device exhibits  well-developed quantum Hall effect at relatively low magnetic fields, demonstrating that our tBLG sample synthesized by CVD is of good quality. The resistances across the SLG-tBLG junction (cross-junction resistances) on the two sides of the Hall bar are measured in details at different magnetic fields. We observe that in the quantum Hall regime, one of the two  cross-junction resistances exhibits a series of quantized plateaus, while the other one tends to stay at zero. These intriguing cross-junction transport properties are well explained by assuming the edge state transport along the Hall-bar boundaries and the junction  interface, based on the multiterminal Landauer-B\"uttker formalism, and thus provide a new physical insight into the edge state transport along the junction interface. We further show that the cross-junction resistance offers a sensitive measure in probing the degeneracy breaking of the Landau levels in tBLG in the quantum Hall regime.

\noindent
\textbf{RESULTS AND DISCUSSION}
\par The graphene junction device studied in this work is made from a CVD-grown SLG-tBLG junction on a Si/SiO$_2$ substrate (see Methods). High crystalline quality graphene films consisting of SLG, bilayer graphene, and as-formed their junctions were grown via CVD (see Supporting Material for details). To evaluate the quality of the as-grown graphene films, Raman spectroscopy measurements were conducted. The left panel of Figure 1a presents a representative Raman spectrum of a graphene film, consisting of SLG, tBLG, and AB-stacked bilayer graphene regions, and the inset of Figure 1a shows a corresponding optical image of the sample. Under the illumination of a 514-nm wavelength laser, the tBLG region exhibits a $\sim$20-fold enhancement in intensity of the Raman G-band compared with the that of the 2D-band. By contrast, the intensities of the Raman G-bands in the SLG region and the AB-stacked bilayer graphene region are rather weak in comparison to their corresponding 2D-bands in the regions. As a fingerprint of tBLG, the resonant enhancement in intensity of the Raman G-band, which has been previously observed in tBLG by several groups\cite{Havener2012,KimK2012,YinJ2016}, is caused by the coupling and interaction between the two layers in tBLG\cite{YinJ2016}. In the Raman mapping graph (the right panel of Figure 1a), the enhanced G-band intensity is uniform over the whole tBLG region, showing an evidence that our tBLG is of high quality. As a comparison, the G-band signature is fairly weak in the AB-stacked bilayer graphene region and becomes nearly invisible in the SLG region. In addition, the clear moir\'e pattern seen in high-resolution transmission electron microscope (HRTEM) image (Figure 1b) further confirms the high crystalline quality of our tBLG samples.

For transport measurements, large-sized graphene films consisting SLG and tBLG regions were transferred onto a heavily doped Si/SiO$_2$ substrates (300 nm in thickness of SiO$_2$). The tBLG regions in the films were identified with the help of optical microscopy and Raman spectroscopy measurements. After selected areas in the film containing SLG-tBLG junctions were patterned into multiterminal Hall-bars by electron beam lithography (EBL) and reactive ion etching, Pd/Au (5 nm/90 nm in thickness) contact electrodes were fabricated via a second step of EBL and metal deposition by electron-beam evaporation.
All electrical measurements were carried out in a Physical Property Measurement System (Quantum Design DynaCool) equipped with a superconducting magnet providing a magnetic field of up to 9 T. The magnetic field was applied perpendicular to the graphene plane.

A sketch for a representative SLG-tBLG junction device and the measurement circuit setup is shown in the top panel of Figure 1c, in which the single layer-bilayer boundary stretches across the middle of the Hall bar.
The multiterminal configuration of the device allows us to characterize the single layer and twisted bilayer region independently.
The voltage probes across the junction, $V_U$ and $V_D$, at the two sides of the sample, are designed to detect the edge state transport along the interface of the SLG-tBLG junction, as we will discuss below.
To further confirm that the bilayer region of the junction is twisted, Raman spatial mapping for a typical SLG-tBLG junction device was carried out and the result is shown in Figure 1c (left bottom panel), in which a uniform Raman G-band intensity over the entire tBLG region is seen. The two right bottom panels of Figure 1c display the Raman spectroscopy curves taken in the SLG and tBLG regions, respectively.
From the low-temperature transfer characteristic curves (Figure S2a), we can estimate, using the nonlinear fitting method\cite{KimS2009}, that the field-effect carrier mobility in the tBLG sample is $\sim$36465 cm$^2$V$^{-1}$s$^{-1}$.
To the best of our knowledge, this value of the mobility is the best result so far reported for the tBLG devices made from CVD-grown tBLG on Si/SiO$_2$ substrates\cite{Fallahazad2012,LuCC2013}.
The quantum Hall plateaus are seen to be well developed at relatively low magnetic fields (seen in Figure S2b), which further confirms the high mobility of the tBLG sample.


Our studies first address the characteristic quantum Hall states in tBLG through magnetotransport measurements performed in the tBLG region at $T =$ 1.9 K.
The Hall resistances $R_{xy}$ measured for the tBLG region as a function of gate voltage $V_{g}$ at $T =$ 1.9 K and at different magnetic fields are displayed in Figure 2a.
We can observe the well-developed plateaus at magnetic fields below 4 T with the Hall resistance following the sequence of 1/$R_{xy}$=$\nu$(e$^2/h$) ($\nu = nh/eB$, $\nu$=4, 12, 20......). The difference in the Hall resistance, 8e$^2/h$, between adjacent well-developed plateaus points to the eight-fold degeneracy of the Landau levels (LLs), originating from the degeneracies of spin (up,down) and valley(K,K'), which can be found in single layer graphene, and additional layer degrees of freedom\cite{SanchezYamagishiJD2012, KimY2013}.
At  magnetic fields beyond 5 T, the new plateaus corresponding to filling factor $\nu$=$\pm 8$ occur (see the dotted lines in Figure 2a). This result indicates that the eight-fold degeneracy of the LLs is lifted by the perpendicular magnetic field. At a high magnetic field, the quantum Hall behavior in tBLG is distinctly different from that conventionally observed in AB-stacked bilayer graphene\cite{Novoselov2006} and in SLG\cite{ZhangY2005,Novoselov2005}.
As schematically shown in the inset of Figure 2a, with increasing magnetic field, the degeneracy of the LLs is lifted and LLs (green lines) arise between the existing ones (blue lines). The twofold layer degeneracy is broken when the eight-fold LLs go beyond the van Hove singularity of tBLG\cite{DeGailR2011,Rozhkov2016}. A better understanding of the LL energy spectrum can be obtained from the Landau fan diagram.
A 2D color plot of the longitudinal resistance $R_{xx}$ as a function of $V_g$ and the magnetic field $B$ is shown in Figure 2b.
We can clearly see that the quantum Hall plateaus for $\nu$=$\pm$4 and $\nu$=$\pm$12 emerge first and, when the magnetic field goes higher, the split of the first excited Landau Level leads to the appearance of filling factor $\nu$=$\pm 8$. The broken degeneracy of  the first excited LL can be observed both on the electron and hole sides. It should be noted that the development of $\nu$=$\pm 8$ in the fan diagram is not the same on the electron and hole sides. One explanation for the electron-hole asymmetry is due to the presence of difference in low-energy van Hove singularity in the hole and electron densities of states of tBLG.


To investigate the impact of the junction interface on the edge states, the longitudinal resistance across the junction (cross-junction resistance) at strong magnetic fields are measured together with the Hall resistances of the SLG and the tBLG regions that constitute the junction. Compared with the pure $R_{xx}$ signals measured in the SLG and tBLG regions, the cross-junction resistances $R_U$ and $R_D$ (see the definitions for $R_U$ and $R_D$ in the schematics shown Figure 3a) can be used to extract the information about the transmission and reflection of edges channels at the interface. Figure 3b shows the measured cross-junction resistance $R_U$ and the Hall resistances $R_{xy}$ of the SLG and tBLG regions against the gate voltage $V_g$ at magnetic field $B=8$ T. It is seen that $R_U$ shows two well-defined quantized plateaus at resistances of $(1/4)h/e^2$ (marked by a green vertical stripe) and $(1/12)h/e^2$ (marked by an orange vertical stripe). The $(1/4)h/e^2$ plateau appears at gate voltages where the SLG Hall resistance is at the $\nu=2$ plateau and the tBLG Hall resistance is at the $\nu=4$ plateau. The cross-junction $(1/12)h/e^2$ plateau appears at gate voltages where the SLG Hall resistance is at the $\nu=-6$ plateau and the tBLG Hall resistance is at the $\nu=-8$ plateau. Here, it should be noted that the filling factors in the SLG and tBLG regions at the gate voltages where the cross-junction resistance stays at quantized plateaus are of integer numbers but can be different. Thus, the quantized cross-junction resistance plateaus are closely related to the  edge states transport (more concretely, i.e., the transmission and reflection of the edge states) at the junction interface.\cite{Williams2007, Haug1988, Washburn1988}.
Since the $V_{dirac}$ in the SLG and tBLG regions nearly coincide in the device, the transport is basically carried out by the same type of carriers in the two region at a given gate voltage.
In the case where the SLG and tBLG regions are in the quantum Hall regime with different Landau-level filling factors, the imbalance in number of the edge channels in the two region will lead to the edge channel reflection at the cross-junction interface. It is this edge channel reflection that results in the observation of the cross-junction quantized resistance plateaus. Considering the type of carriers and the relative relation of filling factors in the SLG and tBLG regions, there are four different situations of the edge channel transport in the SLG-tBLG junction device in the quantum Hall regime, as illustrated in Figure 3a, where the integer filling factors in the SLG and tBLG regions are labeled as $\nu_1$ and $\nu_2$, respectively. The value of the cross-junction resistance measured on the one side of the Hall-bar device $R_U$($R_D$) is quantized, while the cross-junction resistance measured on the other side $R_D$($R_U$) goes to zero. The expected cross-junction quantized resistance in the quantum Hall regime can be derived, based on the Landauer-B\"uttiker formalism\cite{Haug1988, Washburn1988}, as
\begin{equation}
{R}_{cross-junction}={R}_{K}\frac{\left|{\nu}_{1} -{\nu}_{2}\right|}{\left|{\nu}_{1}{\nu}_{2}\right|}                                                                                                          \qquad ({R}_{K}={h/e}^{2})
\label{LB}
\end{equation}
For example, in the range of gate voltages marked by the green vertical stripe in Figure 3b, where $\nu_1$=2 and $\nu_2$=4 (as in the situation described by the lowest panel of Figure 3a), the cross-junction resistance would take a plateau value of $(1/4)h/e^2$ according to Eq.~(1). This is exactly the value  observed in our experiment. In the same way, in the gate voltage region marked by the orange vertical stripe (a hole transport region), where $\nu_1=-6$ and $\nu_2=-4$ (as in the situation described by the second  panel of Figure 3a), the cross-junction resistance would take a plateau value of $(1/12)h/e^2$, in good agreement with our observation. (The same edge channel transport phenomena were also been observed in an SLG-tBLG p-n junction device; for details, see Supporting Materials.)

Figure 3c shows the cross-junction resistances on the upper and lower sides of the Hall-bar device, $R_U$ and $R_D$, at the magnetic fields of $\pm$9 T. Consider first the measurements at the positive magnetic field of $+9$ T (upper panel in Figure 3c). At gate voltages around $V_g=-0.5$ V, the filling factors in the SLG and tBLG regions at this magnetic field are $\nu_1 =-2$ and $\nu_2=-4$. Thus, the measured $R_D$ shows a quantized resistance of (1/4)$h/e^2$, as expected from Eq.~(1), and the measured $R_U$ goes to zero in this gate voltage region. In contrast, $R_U$ is seen to show a quantized conductance of (1/4)$h/e^2$ at gate voltages around $V_g=17.5$ V and a quantized conductance of (1/12)$h/e^2$ at gate voltages around $V_g =-7.4$ V, where $R_D$ goes to zero. These results also agree well with Eq.~(1) derived based on the multiterminal Landauer-B\"uttiker formalism. However, it should be noted that a plateau structure is observed in $R_D$ at gate voltages around at $V_g=24.1$ V and the resistance value at the plateau could not be explained by Eq.~(1). This is because the quantum Hall plateau in the SLG region of our device in this gate voltage region is not fully developed (see Supporting Materials, Figure S3) and, thus, the bulk transport is present in the SLG region which leads to addition of a finite resistance in the cross-junction resistances, $R_D$ and $R_U$, measured in our device. Approximately, this finite resistance contribution is the same in $R_D$ and $R_U$. Thus, the contribution could be subtracted out of the measurements by examining the difference between the measured $R_D$ and $R_U$, as we will discuss below.

The lower panel in Figure 3c shows the same measurements as in the upper panel, but at the reversed magnetic field of $-9$ T. It is seen that the measured $R_U$ ($R_D$) at this negative field resembles well the measured $R_D$ ($R_U$) at the positive field. These results satisfy the Onsager relation which shows that the measured longitudinal resistance (or voltage drop) using the two probes on one side of a symmetric Hall-bar device at a positive magnetic field is equal to the measured longitudinal resistance using the two probes on the other side of the Hall-bar device at the reversed field. Visible deviations between the measured $R_U$ at the positive magnetic field and the measured $R_D$ at the negative field are due to presence of inevitable asymmetry in our device.

In Figure 4a, we plot $|R_U-R_D|$, the absolute value of the difference between the measured $R_U$ and $R_D$, as a function of $V_g$ at $B=9$ T and $T=$ 1.9 K.
In contrast with the measured cross-junction resistances shown in Figure 3c, more plateau-like resistance values can be identified in $|R_U-R_D|$. These resistance values stay closer to the expected quantized values from Eq.~(1) (indicated by the dashed lines) with the corresponding filling factors in the SLG and tBLG regions, $\nu_1$ and $\nu_2$,  assigned and indicated as ($\nu_1,\nu_2$). This better agreement is because as we discussed above, when the quantum Hall plateau is not well developed in a region, a contribution from the bulk transport in the region can be present in the measured individual cross-junction resistances and this contribution could be subtracted out from the measurements by examining $|R_U-R_D|$.
Therefore, the results demonstrate that $|R_U-R_D|$ can be used to identify the edge state transport characteristics in a SLG-tBLG junction device better than individual single-side cross-junction resistances.

Figure 4b shows the gate voltage dependent measurements of $|R_U-R_D|$ at a series of magnetic fields.
Here, again, the filling factors in the SLG and tBLG regions, $\nu_1$ and $\nu_2$, are indicated as ($\nu_1, \nu_2$).
It can be seen that the plateaus corresponding to ($2,4$) and ($-2, -4$) are well developed already at 2 T and get broader with increasing magnetic field. Other plateaus also follow a similar development trend.
It is worth pointing out that the quantized plateaus corresponding to the filling factors ($\nu_1, \nu_2$)=($6,8$) and ($-6,-8$) become visible at a very low magnetic field of 2 T, as indicated by the dash-dotted lines in Figure 4b, indicating that the $\nu_2=8$ and $\nu_2=-8$ quantum Hall states in the tBLG region are detectable in the cross-junction resistance measurements at relatively low magnetic fields. In contrast, in the Hall resistance measurements,
the quantum Hall plateaus of filling factors $\nu_2=8$ and $-8$ in the tBLG region become visible only when the magnetic field exceeds $\sim 5$ T as shown in Figure 2a.
These result demonstrate that the cross-junction resistance $|R_U-R_D|$ is a sensitive tool of detecting the  degeneracy breaking of the LLs in tBLG.

Figure 4c shows the cross-junction resistance $|R_U-R_D|$ of the device measured at the high magnetic field of B=9 T as a function of gate voltage $V_g$ at different temperatures. Here, the plateau structures marked by arrows A, B, and C correspond to  the device at the filling factors of ($\nu_1, \nu_2$)=($-2,-4$), ($-6,-4$), and ($2,4$), respectively. As the temperature increases, the values of $|R_U-R_D|$ on these plateaus move gradually towards smaller values.
Figure 4d show a logarithmic plot of these cross-junction resistance values as a function of temperature. It is seen that at sufficiently high temperatures, the value of $|R_U-R_D|$ on each plateaus follows a thermal activation temperature dependence of $\exp(-kT/E)$.\cite{SchmidtH2013} By line fits of the data in Figure 4d to this thermal activation model in the high temperature range, we obtain the energy scales of activation for the device at filling factors ($-2,-4$), ($-6,-4$), and ($2,4$) as $E_A$=12.6 meV, $E_B$=5.9 meV, and $E_C$=20.1 meV. At filling factors ($-2,-4$) (i.e., on plateau A), holes in the extra $\nu_2=-4$  edge channels in the tBLG region are reflected and turn to transport along the SLG-tBLG junction interface (see the third panel of Figure 2a). A decrease in the cross-junction resistance $|R_U-R_D|$ with increasing temperature arises from a increase in the longitudinal resistance of the carrier transport along the SLG-tBLG junction interface which is dominantly due to thermal activation induced transverse transport through the n=0 and n=1 bulk Landau levels in the tBLG region. Thus, the energy scale $E_A=12.6$ meV corresponds to a measure in the energy difference between the zero energy and the first hole Landau levels in the tBLG region at B=9 T. In contrast, at filling factors ($-6,-4$) (i.e., on plateau B), holes in the extra $\nu_1=-6$  edge channels in the SLG region are reflected and turn to transport along the SLG-tBLG junction interface (see the second panel of Figure 2a). As a results, the energy scale $E_B=5.9$ meV corresponds to a measure in the energy difference between the first and second hole Landau levels in the SLG region at B=9 T. The fact that a smaller value of $E_B$ than $E_A$ is observed is in a good consistence with the theoretical prediction that the energy spacing between two adjacent Landau levels decreases with increasing Landau level index\cite{Sarma2011}. At filling factors ($2,4$) (i.e., on plateau C), the system is in the electron transport regime. Here, electrons in the extra $\nu_2=4$ edge channels in the tBLG region are reflected and turn to transport along the SLG-tBLG junction interface (see the last panel of Figure 2a). Thus, the energy scale $E_C=20.1$ meV presents a measure of the energy difference in n=0 and n=1 electron Landau levels in the tBLG region at B=9 T. The difference seen between $E_A$ and $E_C$ reflects the presence of the electron-hole asymmetry in the tBLG film.

\noindent
\textbf{CONCLUSION}

In summary, we have fabricated a SLG-tBLG junction device in a Hall-bar setup and have studied the transport properties of the device.
In the quantum Hall regime, we have observed a series of quantized plateaus in the longitudinal resistance measured across the SLG-tBLG junction of the device. We have showed that the resistance values on these plateaus can be described using the Landauer-B\"uttiker formalism and demonstrated the appearance of the edge channel transport along the SLG-tBLG junction interface in the device.
We have also showed that the difference between the cross-junction resistances measured on the two sides of the Hall-bar provides a sensitive measure to edge-channel transport characteristics in the two graphene layers that constitute the junction and to degeneracy lifting of the Landau levels in the tBLG layer. Temperature dependent measurements of the cross-junction resistance have also been carried out and the influence of the transverse transport of the bulk Landau levels on the edge channel transport along the junction interface has been extracted.
Our results reveal lateral graphene hybrid junction structures exhibit rich transport physics and measurements of the cross-junction transport properties in such junction structures open up new opportunities for probing exotic quantum phenomena in graphene devices.

\noindent
\textbf{METHODS}

\noindent
\textbf{tBLG growth and characterization}.
The twisted bilayer graphene (tBLG) sample used in this study was synthesized on a Cu foil in a low pressure CVD system. First, the Cu foil (Alfa-Aesar \#46365) was annealed in a quartz tube furnace (Lindberg/Blue M) at 1020 $^{\circ}$C under a 500 standard cubic centimeters per minute (sccm) flow of H$_2$ for 1 hour. The tBLG sample was then grown under a 1000 sccm flow of H$_2$ and a 0.8 sccm flow of CH$_4$ for 40 min at 910 Pa. After transferring onto a Si/SiO$_2$ substrate, the grown tBLG sample was examined using an optical microscope (Olympus BX51 and Nikon LV100ND) and by Raman spectroscopy (Horiba HR800, 100 objective, under a wavelength of 514 nm) measurements. The moir\'e pattern image was taken using an aberration-corrected TEM (FEI Titan Cubed Themis G2 300 under 300 kV).

\noindent
\textbf{Graphene transfer}.
The grown tBLG sample was transferred onto a Si/SiO$_2$ substrate (with 300-nm-thick SiO$_2$)  for device fabrication by a poly(methyl methacrylate) (PMMA)-assisted technique. First, the graphene on Cu foil sample was spin-coated with PMMA at 1000 rpm (revolutions per minute) on one side and was subsequent baked at 170 $^{\circ}$C for 5 min.  The graphene layer on the uncoated side of the sample was then etched by air plasma (with flux of 15 sccm and power of 90 W) for 2 min. The as-formed PMMA/graphene/Cu sandwich structure was placed and floated on 1 M Na$_2$S$_2$O$_8$ etching solution. After etching the Cu foil away, the freestanding PMMA/graphene film was rinsed four times in deionized water, and was then washed by isopropanol for 30 s and dried in clean air for 12 h. Subsequently, The PMMA/graphene film was placed onto the Si/SiO$_2$ substrate. Finally, the PMMA was dissolved with acetone, leaving the tBLG layer on the Si/SiO$_2$ substrate.

\noindent
\textbf{Device fabrication and transport measurements}.
Subsequent to transferring the tBLG sample onto the Si substrate with 300 nm-thick SiO$_2$ on top, the flatness of the sample was checked with an atomic force microscope (AFM). Electron beam lithography (EBL) (Raith 150 II) and reactive ion etching (RIE) (Trion technology minilock III) were employed to pattern the graphene sample into a Hall bar. Using EBL and electron-beam evaporation (Kurte J. Lesker AXXIS), palladium/gold (5/90 nm  in thickness) electrodes were made to the graphene sample. Transport measurements were performed at temperatures of 300 K to 1.9 K and magnetic fields of up to 9 T in a Physical Property Measurement System (Quantum Design, DynaCool). The electrical signals were measured by Stanford SR830, using a current-biased lock-in technique at a frequency of 17.777 Hz.

\newpage

\newpage
\noindent
\textbf{Author Information}
\noindent
\par
\textbf{Corresponding Authors}
\par *Email: nkang@pku.edu.cn
\par *Email: hqxu@pku.edu.cn
\par

\textbf{Notes}
\par The authors declare no competing financial interest.

\noindent
\textbf{Acknowledgements}

We thank the Electron Microscopy Laboratory, School of Physics, Peking University for the help on high-resolution TEM characterization. We acknowledge financial supports by the Ministry of Science and Technology of China through the National Key Research and Development Program
of China (Grant Nos. 2016YFA0300601 and 2017YFA0303304), the National Natural Science Foundation of China (Grant Nos. 11774005, 11874071, and 11974026), and the Beijing Academy of Quantum Information Sciences (Grant No. Y18G22). \\

\noindent{\bf References}


\clearpage

\begin{figure}
\begin{center}
\includegraphics[width=5.5in]{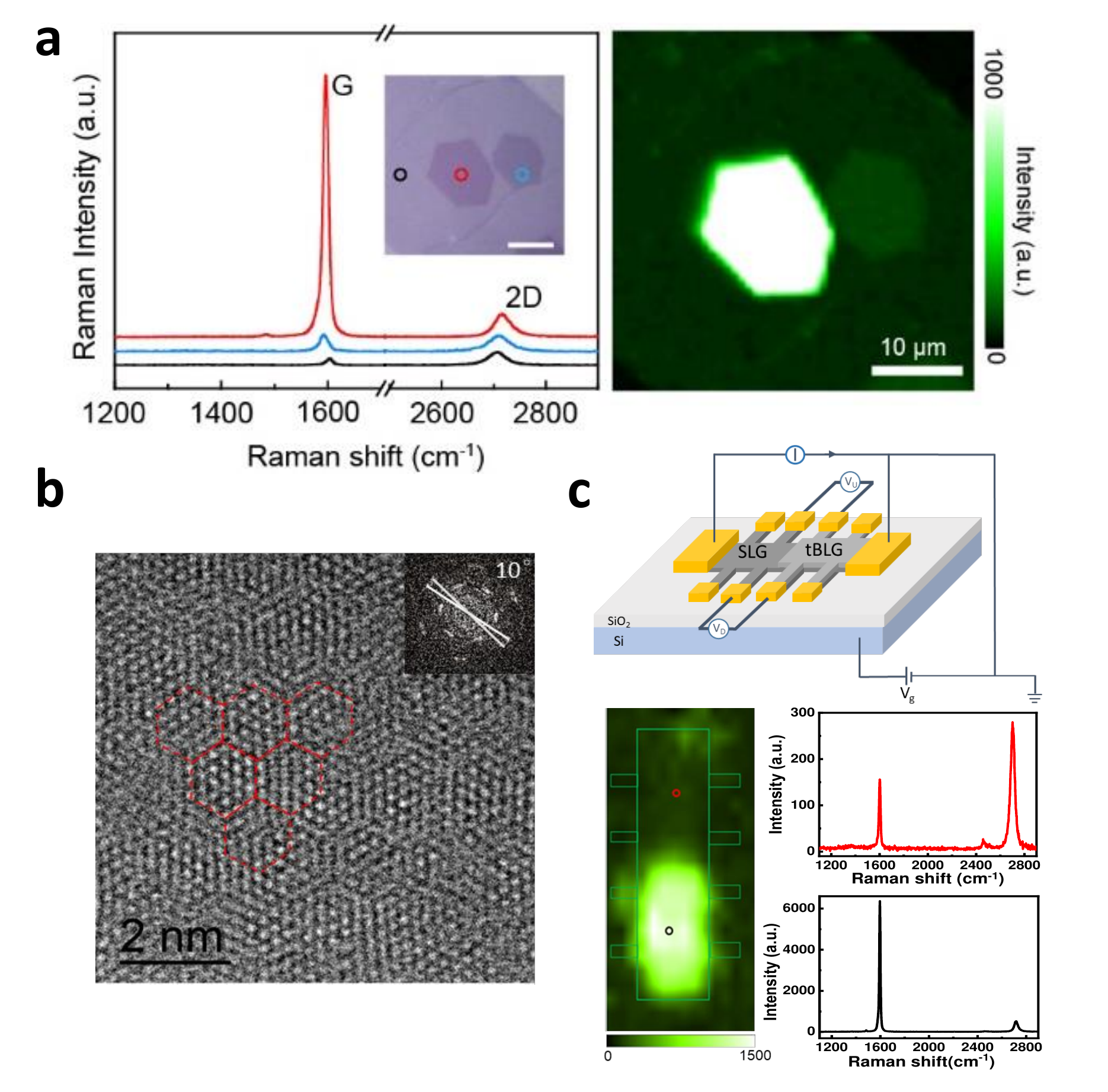}
\caption{
\textbf{a.} The left panel shows the Raman spectra of SLG (black curve), AB-stacked bilayer graphene (blue curve) and tBLG (red curve) region taken under an incident laser of 514 nm in wavelength (2.41 eV). The inset shows an optical image of a graphene sample, in which the black, blue and red circle mark the regions containing SLG, AB-stacked bilayer graphene and tBLG with a $13^{\circ}$ twisted angle, on a SiO$_2$ substrate. Here, the scale bar corresponds to 10 $\mu$m.
The right panel shows a G-band intensity mapping image of the tBLG region\cite{KimK2012}. Here a good uniformity of the intensity enhancement of Raman G-band is seen.
\textbf{b.} High-resolution TEM image of a tBLG sample. The scale bar corresponds to 2 nm. The inset is the fast Fourier transform (FFT) of the image, showing that the twisted angle in this sample is $\sim 10^{\circ}$.
\textbf{c.} Top: sketch of a SLG-tBLG junction device. Bottom left: Raman G-band intensity spatial mapping image of a SLG-tBLG junction device, in which tBLG has a twisted angle of $\sim 10^{\circ}$.\cite{KimK2012} The green line marks the Hall bar structure of the device. The red and black circles mark, respectively, the SLG and tBLG regions where the Raman spectra are taken. Bottom right: Raman spectra taken in the SLG region marked by the red circle in the bottom left panel (red curve) and in the tBLG region marked by the black circle in the bottom left panel (black curve). Here, the incident laser wavelength is set at 633 nm (1.96 eV).
}
 \label{fig1}
\end{center}
\end{figure}

\clearpage

\begin{figure}
\begin{center}
\includegraphics[width=6.0in]{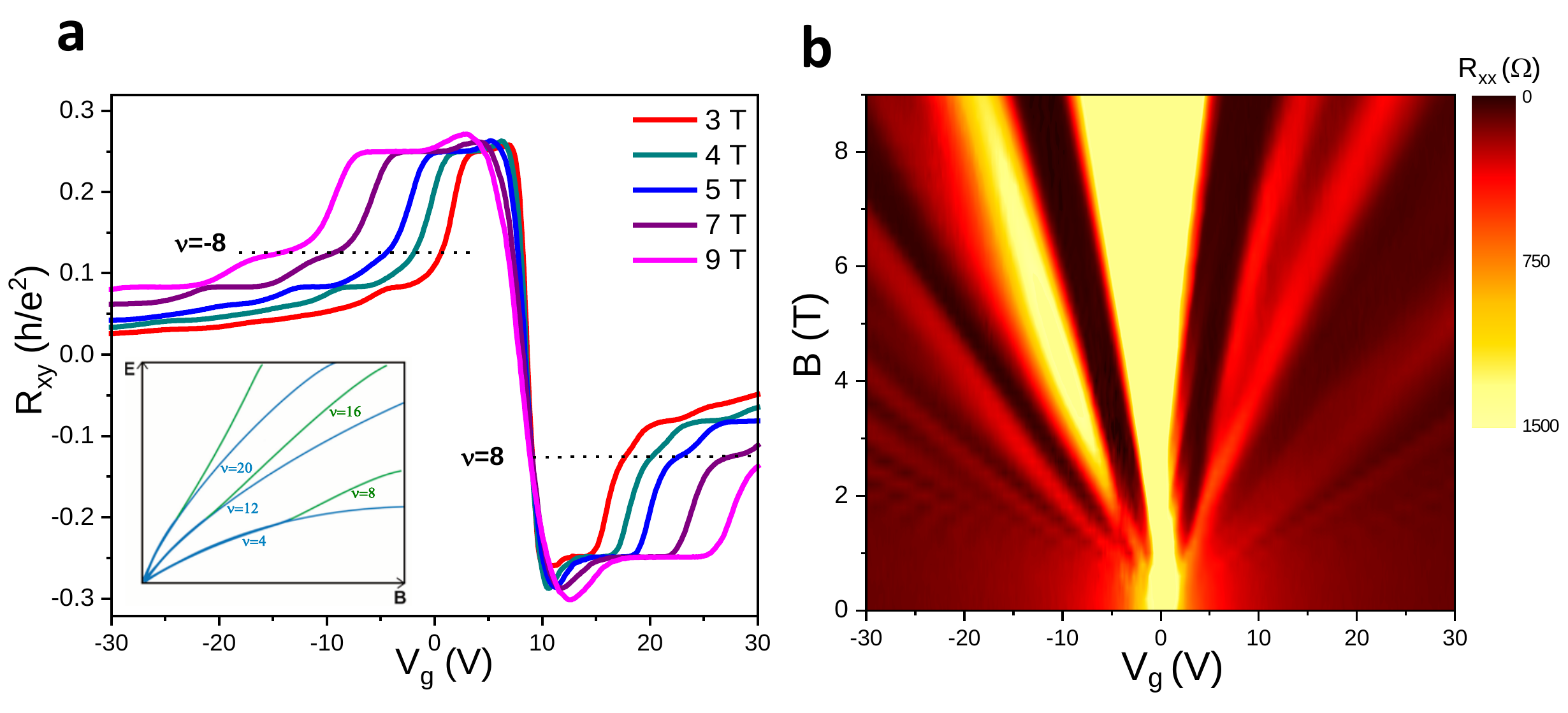}
\caption{
\textbf{a.} Hall resistance of tBLG measured for the device as shown in the bottom left panel of Figure 1c as a function of gate voltage $V_{g}$ at $T =$ 1.9 K and different magnetic fields. The inset shows a schematic illustration of the Landau levels (LLs) spectrum of tBLG, in which the values of $\nu$ correspond to the filling factors in the gaps. The light blue lines represent the original eightfold degenerate LLs at low magnetic fields. The green lines show the emerging LLs when the magnetic field becomes large enough to break the eightfold degeneracy of the LLs in tBLG.
\textbf{b.} 2D contour plot of the longitudinal resistances ($R_{xx}$) of tBLG measured for another but similar device as a function of gate voltage and magnetic field at $T =$ 1.9 K.
}
\label{fig2}
\end{center}
\end{figure}

\clearpage

\begin{figure}
\begin{center}
\includegraphics[width=6.0in]{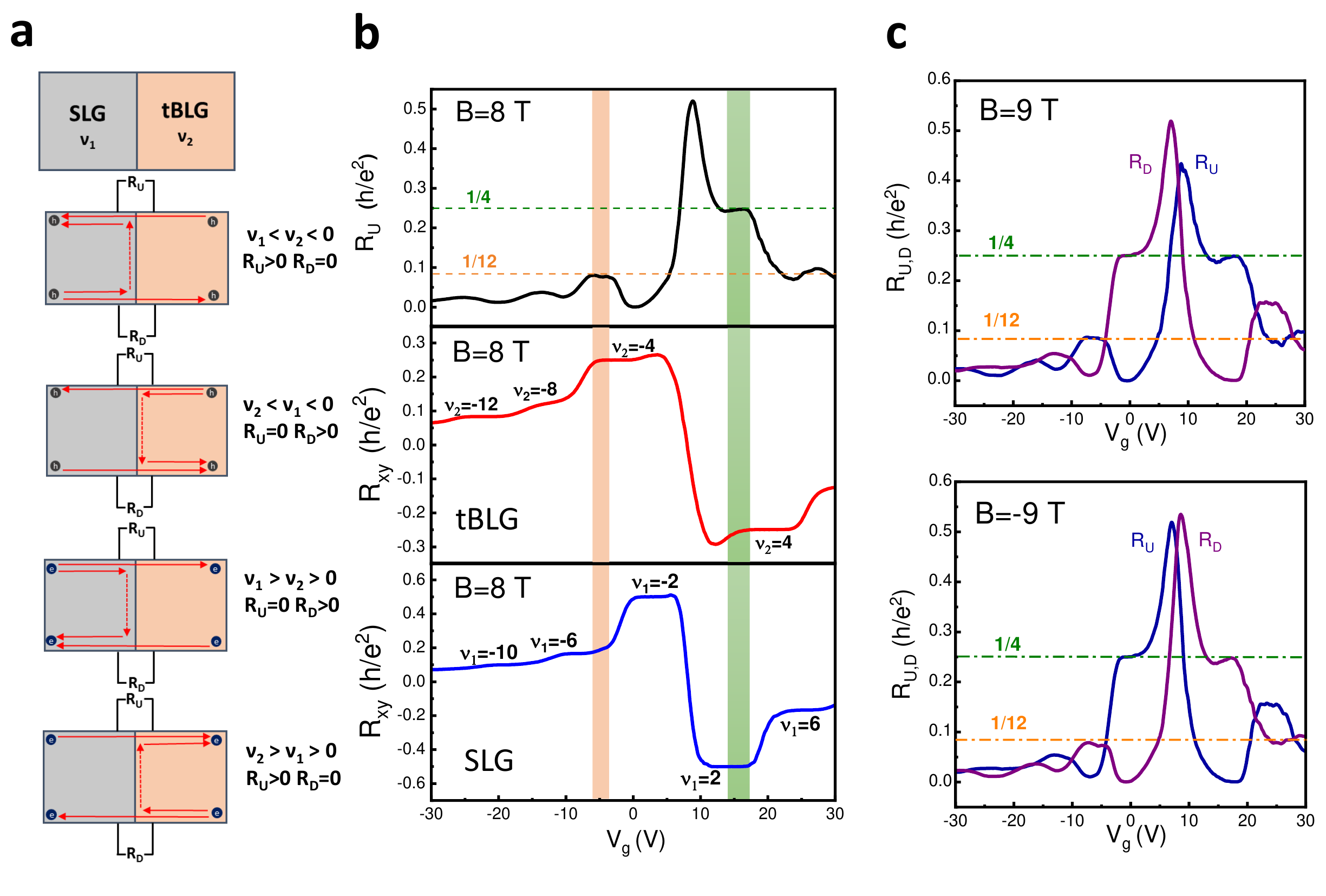}
\caption{
\textbf{a.} Schematic illustration of edge-state transport in the SLG-tBLG junction device in the quantum Hall regime at different carrier configurations, where $\nu_1$ and $\nu_2$ represent the filling factors in the SLG and tBLG regions, respectively.
\textbf{b.} Top: Cross-junction resistance $R_U$ measured for the device shown in the bottom left panel of Figure 1c as a function of gate voltage $V_{g}$ at the perpendicularly applied magnetic field of 8 T.
Central and bottom: Simultaneously measured Hall resistances for the tBLG and SLG regions.
The green region represents a gate voltage range of $\nu_1$=2 in the SLG region and $\nu_2$=4 in the tBLG region, and $R_U = \frac{1}{4}$ $\frac{h}{e^2}$ in the cross-junction resistance.
The orange region represents a gate voltage range of $\nu_1$= -6 in the SLG and $\nu_2$= -4 in the tBLG, and $R_U=\frac{1}{12}$ $\frac{h}{e^2}$ in the cross-junction resistance.
\textbf{c.} Cross-junction resistances $R_U$ and $R_D$ measured for the device at the magnetic fields of $B=\pm$9 T.
}
\label{fig3}
\end{center}
\end{figure}

\clearpage

\begin{figure}
\begin{center}
\includegraphics[width=6.0in]{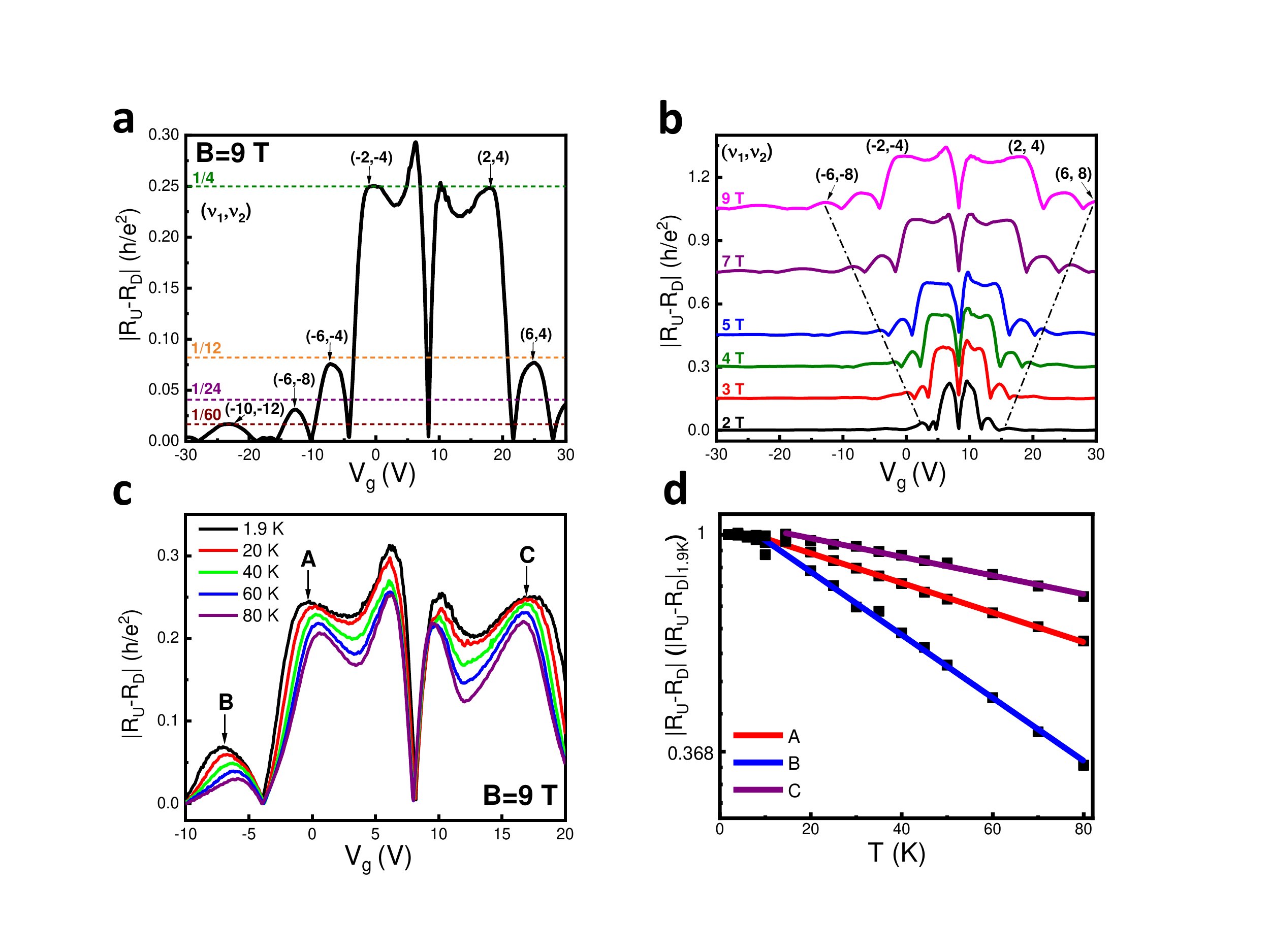}
\caption{
\textbf{a.} Absolute value of the difference between $R_U$ and $R_D$, i.e., $|R_U$-$R_D|$, of the device shown in the bottom left panel of Figure 1c as a function of gate voltage $V_g$ at $B=$9 T and $T =$ 1.9 K. ($\nu_{1}$,$\nu_{2}$) label the filling factors in the SLG and tBLG regions determined from the Hall resistance measurements. The dotted lines mark the expected quantized values of $\frac{1}{4}$$\frac{h}{e^2}$ (green), $\frac{1}{12}$$\frac{h}{e^2}$ (orange), $\frac{1}{24}$$\frac{h}{e^2}$ (purple), and $\frac{1}{60}$$\frac{h}{e^2}$ (brown) in $|R_U$-$R_D|$.
\textbf{b.} $|R_U$-$R_D|$ vs. $V_g$ at magnetic fields of 2 T to 9 T. The plateaus associated with $\nu_2=\pm8$ can be observed at a magnetic field as low as $B=2$ T.
\textbf{c.} $|R_U$-$R_D|$ vs. $V_g$ taken at $B=9$ T and different temperatures.
\textbf{d.} Logarithmic plots of the normalized values of $|R_U$-$R_D|$ as a function of temperature at the gate voltages marked by A, B and C in \textbf{c}, where $|R_U$-$R_D|$ is expected to be on plateaus. Here, the data show good exponential temperature-dependent  characteristics, $\exp(-kT/E)$, except for at very low temperatures.
}
\label{fig4}
\end{center}
\end{figure}

\clearpage

\end{document}